\documentclass{ws-procs10x7square}

\begin{document}


\title{ABOUT THE CAVITY FIELDS IN MEAN FIELD SPIN GLASS MODELS}

\author{Francesco Guerra}
\address{Dipartimento di Fisica, Universit\`a di Roma ``La Sapienza''
\\ and INFN, Sezione di Roma, Piazzale Aldo Moro 2, I-00185 Roma, 
Italy\\e-mail: {\tt francesco.guerra@roma1.infn.it}}  


\maketitle

\abstracts{
We give the explicit expression of the infinite volume limit for the random 
overlap structures appearing in the mean field spin glass model. These 
structures have the expected factorization property for the cavity fields, 
and enjoy invariance with respect to a large class of 
random transformations. These 
properties are typical of the Parisi {\it Ansatz}. 
}


\section{Introduction}

In previous work \cite{GTthermo}, \cite{GTcergy}, by using a simple 
interpolation 
argument, we have proven 
the existence of the thermodynamic limit of the free energy, for mean field disordered models, 
including the Sherrington-Kirkpatrick model \cite{SK}, and the 
Derrida p-spin model \cite{D}. 
Then, we have extended this argument \cite{Grepli}, in order to compare the 
limiting free energy 
with the expression given by the Parisi {\it Ansatz} \cite{MPV}, 
and including full 
spontaneous replica symmetry breaking. Our main result is that the quenched 
average of the free energy is bounded from below by the value given in the Parisi 
{\it Ansatz}, uniformly in the size of the system. Moreover, the difference
between the two expressions is given in the form of a sum rule, extending 
our previous work on the comparison between the true free energy and its 
replica symmetric Sherrington-Kirkpatrick approximation \cite{Gsum}. We 
have given also a 
variational bound for the infinite volume limit of the ground state energy 
per site.

In the meantime, Aizenman, Sims, and Starr \cite{ASS} have been able to 
formulate a 
very attractive extended variational principle, in which the actual value 
of the free energy, in the infinite volume limit, is expressed through an 
optimization procedure for which the Parisi ultrametric/hierarchical 
structures form only a subset of the variational class. This leads the 
authors to put the question whether ``ultrametricity is an inherent structure 
of the Sherrington-Kirkpatrick mean-field model, or is it only a 
simplifying assumption''. 

Finally, Talagrand \cite{Tb} has announced a 
proof, based on a far reaching extension of the methods proposed in 
\cite{Grepli}, of the long expected result that the infinite volume 
limit of the free energy is precisely given by the Parisi expression.

The main purpose of this paper is to report about the general form of 
the random overlap structures arising in the infinite volume limit for the 
cavity fields. We will show that they have a very important factorization 
property, and, moreover, they are invariant under a large class of random 
transformations. These properties are typical of the Parisi {\it Ansatz}.

The organization of the paper is as follows. In Section 2, we will briefly 
recall the main features, with definitions and results, of the mean field spin glass 
model. Section 3 contains the main result of this report. We give only a 
rapid sketch of the proof, by pointing out the main ideas. A more detailed 
treatment will be presented elsewhere \cite{GTultra}. Finally, we give 
also a synthetic  outlook for further developments.       

\section{The basic definitions and results for the mean field spin glass model}

The generic configuration of the mean field spin glass model is defined 
through Ising spin variables
$\sigma_{i}=\pm 1$,  attached to each site $i=1,2,\dots,N$. 

The external quenched disorder is
given by the $N(N-1)/2$ independent and identically distributed random
variables $J_{ij}$, defined for each unordered couple of sites. For the sake of simplicity,
we assume each $J_{ij}$ to be a centered
unit Gaussian with averages $E(J_{ij})=0$, $ E(J_{ij}^2)=1$.

The Hamiltonian of the model, in some external field of strength $h$,  
is given by the mean field expression
\begin{equation}\label{H}
H_N(\sigma,h,J)=-{1\over\sqrt{N}}\sum_{(i,j)}J_{ij}\sigma_i\sigma_j
-h\sum_{i}\sigma_i.
\end{equation}
Here, the first sum extends to all site couples, an the second to all sites.

For a given inverse temperature $\beta$, let us now introduce the 
disorder dependent
partition function $Z_{N}(\beta,h,J)$, 
the quenched average of the free energy per site
$f_{N}(\beta,h)$, the Boltzmann state 
$\omega_N$, and the auxiliary function $\alpha_N(\beta,h)$,  
according to the well known definitions
\begin{eqnarray}\label{Z}
&&Z_N(\beta,h,J)=\sum_{\sigma_1\dots\sigma_N}\exp(-\beta H_N(\sigma,h,J)),\\
\label{f}
&&-\beta f_N(\beta,h)=N^{-1} E\log Z_N(\beta,h,J)=\alpha_N(\beta,h),\\
\label{state}
&&\omega_{N}(A)=Z_N(\beta,h,J)^{-1}\sum_{\sigma_1\dots\sigma_N}A\exp(-\beta
H_N(\sigma,h,J)), 
\end{eqnarray}
where $A$ is a generic function of the $\sigma$'s. Notice that the 
Boltzmann state is depending also on the external noise $J_{ij}$.  

Replicas are introduced by 
considering a generic number $s$ of independent copies
of the system, characterized by the Boltzmann
variables $\sigma^{(1)}_i$, $\sigma^{(2)}_i$, $\dots$,
distributed according to the product state 
$\Omega_N=\omega^{(1)}_N \omega^{(2)}_N \dots\omega^{(s)}_N$. Here,
all $\omega^{(\alpha)}_N$ act on each one
$\sigma^{(\alpha)}_i$'s, and are subject to the {\sl
same} sample $J$ of the external noise. 

The overlap between  two replicas $a,b$ is
defined according to
$q_{ab}=
N^{-1}\sum_{i}\sigma^{(a)}_i\sigma^{(b)}_i,$
with the obvious bounds
$-1\le q_{ab}\le 1$.

For a generic smooth function $F$ of the overlaps, we
define the $\langle\,\,\rangle$ averages
\begin{equation}
\label{medie}
\langle F(q_{12},q_{13},\dots)\rangle=
E\Omega_N\bigl(F(q_{12},q_{13},\dots)\bigr),
\end{equation}
where the Boltzmann averages $\Omega_N$ act on the replicated $\sigma$
variables, and $E$ is the average with respect to the external noise $J$.

In \cite{GTthermo} it was shown how to prove the existence of the limit
$$\lim_{N\to\infty}\alpha_N(\beta,h)=\sup_N \alpha_N(\beta,h)\equiv 
\alpha(\beta,h),$$ for all values of the parameters $(\beta,h)$.

By using standard convexity arguments, it is easy to prove, for generic
values of the parameters $(\beta,h)$, {\it i.e.} by excluding a set of 
zero measure, that the following limit also exists
$$\lim_{N\to\infty}\langle q^{2}_{12} \rangle_N
\equiv \langle q^{2}_{12} \rangle,$$
where we have stressed the dependence on $N$ before the limit.   

Moreover, the results of \cite{Grepli} show that $\alpha(\beta,h)$ is 
uniformly bounded from above by the value found in the frame of the Parisi 
{\it Ansatz}.

\section{The infinite volume limit of the random overlap structure}

A very general definition of random overlap structures has been given in 
\cite{ASS}. Important references dealing with this concept are 
\cite{R}, \cite{N}, \cite{BS}, \cite{T}, \cite{RA}. Here we need a much simpler setting, 
because our random overlap structures will be explicitely constructed 
through an infinite volume limit.

Let us consider the Boltzmann state, as defined in previous 
section, called here  $\Omega_N$,  acting on the $N$ Ising variables 
called here $\tau_1, 
\tau_2,\dots, \tau_N$, for the sake of convenience. Let $\eta_j(\tau)$, 
for $j=1,2,\dots$ be $\tau$-conditionally independent Gaussian random variables, 
with zero mean 
and variance given by the overlaps
\begin{equation}\label{eta}
E(\eta_j(\tau) \eta_{j^\prime}(\tau^\prime)) =
\delta_{j j^\prime} q(\tau,\tau^\prime).
\end{equation}
We introduce also the family of Gaussian random variables, $\kappa(\tau)$, 
as independent from the $\eta$'s, with zero mean 
and variance given by the square of the overlaps
\begin{equation}\label{kappa}
E(\kappa(\tau) \kappa(\tau^\prime)) =
q^{2}(\tau,\tau^\prime).
\end{equation}

We need also the random variables defined by
\begin{equation}\label{c}
c_j=2 \cosh\beta(h+\eta_j)=\sum_{\sigma_j} \exp(\beta(h+\eta_j)\sigma_j.
\end{equation}

Consider also the integer random variable $K$ uniformly distributed on the 
values $(1,2,\dots,N)$.

Now we are ready to define our main quantities. For any integer $M$, and 
any real $\lambda$, introduce
\begin{equation}\label{Omega}
E\log\Omega_K(c_1 c_2 \dots c_M \exp(\lambda\kappa)).
\end{equation}
Here the average $E$ contains the averages over the external noise 
$J_{ij}$, 
appearing in $\Omega_K$, over the variables $\eta$ and $\kappa$, and over 
$K$.

It is simple to realize that the previously introduced expression has a 
very simple interpretation. In fact, we can consider  a large number of 
cavity spins $\tau$ coupled with some additional spins $\sigma$, and 
evaluate the change in free energy coming from the $M$ added spins and from a 
small change in the two-spin coupling in the cavity.

The need to consider the additional average over $K$ is due to the 
uncomplete control of the corrections to the infinite volume limit for the 
free energy. This will amount to the consideration of a kind of Ces\`aro 
limit in the following. Should the corrections be found of order 
$O(\frac{1}{N})$, as naturally expected, then a simple limit over the size 
of the cavity would be sufficient.

The main result of this report is the following.
\begin{theorem}
For all values of the $(\beta,h)$ parameters, where the infinite volume 
limit of the averaged squared overlap $\langle q^{2}_{12}\rangle$ is uniquely 
defined,  we have
\begin{equation}\label{lim}
\lim_{N\to\infty} E\log\Omega_K(c_1 c_2 \dots c_M \exp(\lambda\kappa))=
M(\alpha(\beta,h)+\frac{\beta^2}{4}(1-\langle q^{2}_{12}\rangle))+
\frac{\lambda^2}{2}(1-\langle q^{2}_{12}\rangle).
\end{equation}
\end{theorem}

The proof is long but straigthforward. The main ingredients are given by the 
convergence of the free energy, standard convexity arguments, and the fact 
that for a very large cavity the added spins $\sigma$'s do not interact 
among themselves, but only with the cavity spins $\tau$ through the random 
fields $\eta$. Notice that in the Ces\`aro limit only the large values of 
$K$ do really matter in the limit on $N$. We refer to \cite{GTultra} for all 
details.

Let us call $E\log\Omega$ simbolically the resulting random overlap 
limiting structure. Notice that it shares a very important factorization 
property. In fact, the random variables $\eta,\kappa$ become independent 
under the limiting $E\log\Omega$. As a consequence, it turns out that 
$\Omega$ is invariant under all stochastic transformations of the type
$$\Omega(.) \to \Omega^{\prime}(.) \equiv \Omega(c_1 .)/\Omega(c_1),$$
$$\Omega(.) \to \Omega^{\prime\prime}(.) 
\equiv \Omega(\exp(\lambda\kappa) .)/\Omega(\exp(\lambda\kappa)).$$

These factorization and invariance properties are typical of the random 
overlap structures found in the frame of the Parisi {\it Ansatz}. It is 
also clear that in the extended variational principle of Aizenman, Sims and 
Starr, without loss of generality, we can limit ourselves to the 
consideration only of factorized overlap structures. In this way we get 
always bounds uniform in the size of the system.

The set of general factorized overlap structures enjoys a natural convexity 
property. In fact, if $\Omega_i$, $i=1,2$, are factorized, then the new 
structure, defined by $E\log\Omega_i$, where $E$ contains also a convex 
average over $i$, is clearly factorized. This shows that the set of all 
factorized structures is larger then the Parisi set. In fact, the mixture 
of two Parisi structures is not Parisi, in general. However, only the 
extremal structures are relevant in the optimization procedure in the 
extended variational principle. Therefore, the problem to see whether the
optimal structure is ultrametric or not is still open. We plan to dedicate 
future work to this important problem.

Finally, let us remark that there are strong connections between our 
general results about limiting random overlap structures, and the 
stability properties found in the frame of the cavity approach of 
Mezard-Parisi-Virasoro \cite{MPVcavity}, or the kind of stochastic stability
exploited by Aizenman and Contucci in \cite{AC}.



\section*{Acknowledgments}

We gratefully acknowledge useful conversations with Michael Aizenman, 
Pierluigi Contucci,
Giorgio Parisi and Michel Talagrand. The strategy developed in this paper 
grew out from a 
systematic exploration of interpolation methods, developed in 
collaboration with Fabio Lucio Toninelli.

This work was supported in part by MIUR 
(Italian Minister of Instruction, University and Research), 
and by INFN (Italian National Institute for Nuclear Physics).



\end{document}